\documentstyle[12pt]{article}
\begin{document}

\begin{titlepage}
\normalsize
\begin{center}
{\Large \bf Budker Institute of Nuclear Physics} 
\end{center}
\begin{flushright}
BINP 96-3\\
January 1996
\end{flushright}
\vspace{1.0cm}
\begin{center}
{\bf GRAVITATIONAL INTERACTION OF SPINNING BODIES,}
\end{center}
\begin{center}
{\bf CENTER-OF-MASS COORDINATE}
\end{center}
\begin{center}
{\bf AND RADIATION OF COMPACT BINARY SYSTEMS}
\end{center}

\vspace{1.0cm}
\begin{center}
{\bf I.B. Khriplovich}\footnote{e-mail address: khriplovich@inp.nsk.su}
{\bf and A.A. Pomeransky}\footnote{Novosibirsk University}
\end{center}

\begin{center}
Budker Institute of Nuclear Physics, 630090 Novosibirsk,
Russia
\vspace{3.0cm}
\end{center}

\begin{abstract}
Spin-orbit and spin-spin effects in the gravitational interaction 
are treated in a close analogy with the fine and hyperfine interactions
in atoms. The proper definition of the cener-of-mass coordinate is
discussed. The technique developed is applied then to the gravitational
radiation of compact binary stars. Our result for the spin-orbit
correction differs from that obtained by other authors. New effects
possible for the motion of a spinning particle in a gravitational field
are pointed out. The corresponding corrections, nonlinear in spin, are
in principle of the same order of magnitude as the ordinary spin-spin
interaction.
\end{abstract}

\end{titlepage}

{\bf 1.} It is expected that in few years the gravitational radiation
from coalescing binary stars will be observed by laser interferometer
systems LIGO and VIRGO. Its successful detection depends crucially
on the accurate theoretical prediction of the exact form of the signal.
In this way the observed effect becomes sensitive to the relativistic
corrections of the $c^{-2}$, $c^{-3}$ and $c^{-4}$ orders to the motion
of a binary system and to the radiation intensity. In  particular, the
spin-orbit interaction becomes essential, and for two extreme Kerr
black holes even the spin-spin one \cite{kww}.

Some years ago it was noticed that the general relativity can accomodate
in a natural way a specific gravitational magnetic moment coupling
\cite{kh} (see also \cite{yb}). The starting point of the present work
was the observation that the spin self-interaction arising in this way
is of the same order of magnitude as the spin-spin interaction, and
therefore in principle its existence can be checked in the
gravitational-wave experiments. 

However, in the course of the investigation, when trying to rederive
previous calculations related to the spin effects in the gravitational
radiation of binary stars, we came to the results which differ from
those of Refs. \cite{kww,bdi} as concerns the spin-orbit contributions.
The origin of this discrepancy can be traced back to, what is to our
belief, a long-standing confusion concerning the definition of the
centre of mass in the case when spin is taken into account. The problem
is quite instructive and amusing by itself, and on the other hand, the
spin-orbit correction is the leading one among spin effects. That is why
we would like to start our discussion with this subject.

\bigskip

{\bf 2.} The spin-orbit and spin-spin interactions in the two-body
problem can be immediately obtained in fact from the well-known results
for the limiting case when one of the bodies (say, 2) is very heavy
(see, e.g., book \cite{ll}). In this limit we have the usual spin-orbit
interaction \cite{ds}
\begin{equation}\label{1ls1}
V^1_{1ls}=\,\frac{3}{2}\,\frac{k}{c^2 r^3}\,\frac{m_2}{m_1}\,\vec l
\vec s_1,
\end{equation}
the interaction of the orbital angular momentum $\vec l$ with the spin
$\vec s_2$ of the central body \cite{tl}
\begin{equation}
V^1_{2ls}=\,2\,\frac{k}{c^2 r^3}\,\vec l \vec s_2,
\end{equation}
and the spin-spin interaction \cite{sch}
\begin{equation}\label{ss}
V_{ss}=\,\frac{k}{c^2 r^3}\,[3(\vec s_1 \vec n) (\vec s_2 \vec n) -
\vec s_1 \vec s_2]. 
\end{equation}
Simple symmetry arguments dictate  now the form of the spin-orbit 
interaction for the two-body problem:
\begin{equation}\label{1ls}
V_{1ls}=\,\frac{3}{2}\,\frac{k}{c^2 r^3}\,\vec l \left(\frac{m_2}{m_1}\,
\vec s_1+\,\frac{m_1}{m_2}\,\vec s_2\right),
\end{equation}
\begin{equation}\label{2ls}
V_{2ls}=\,2\,\frac{k}{c^2 r^3}\,\vec l(\vec s_1 +\,\vec s_2).
\end{equation}
As to the spin-spin interaction, it is of the same form (\ref{ss}).

However, due to the mentioned discrepancy concerning the spin-orbit
corrections to the gravitational radiation, it turns out expedient to
derive explicitly the interactions discussed. This is only an elementary
generalization of the solution of the corresponding problem for the case
of a heavy central body, as given in book \cite{ll} (\S 106, Problem 4).
We will start with the two-body Lagrangian including
$c^{-2}$ corrections:
$$L=\,\frac{m_1 v_1^2}{2}\, +\, \frac{m_2 v_2^2}{2}\, +\,\frac{k m_1
m_2}{r}\,+\,\frac{m_1 v_1^4 + m_2 v_2^4}{8 c^2}\,$$
\begin{equation}\label{L}
 +\,\frac{k m_1 m_2}{2
c^2 r}\,[3(v_1^2\,+\,v_2^2)\,-\,7(\vec{v}_1\vec{v}_2)-\,
(\vec{v}_1\vec{n})(\vec{v}_2\vec{n})]-\,\frac{k^2 m_1 m_2 (m_1+m_2)}{2
c^2 r^2}.
\end{equation}
Here $\vec{r}=\vec{r}_1-\vec{r}_2;\;\;\vec{n}=\vec{r}/r;\;\;m_i,\;
\vec{r}_i,\; \vec{v}_i$ are the mass, coordinate and velocity,
respectively, of the $i$th particle, $i=1,2$. 

Let us take the term with $v_1^2$ in the second line of Eq. (\ref{L}).
We write the velocities of individual elements of the top 1 (with mass
$dm_1$) in the form
$$\vec{v}_1+\vec{\omega}_1\times \vec{\rho}_1,$$ 
where $\vec{v}_1$ is the velocity of the orbital motion, 
$\vec{\omega}_1$ is the angular velocity. The radius-vector
$\vec{\rho}_1$ of the element $dm_1$ is counted off the center of mass of
the top 1, so that the integral over the volume of the top 
\begin{equation}\label{cm}
\int \vec{\rho}_1 dm_1\,=0.
\end{equation}
Due to (\ref{cm}) the first-order term of the expansion in $\rho_1/r$ of
the interaction discussed vanishes. As to the second-order term, with
the obvious definition 
$$\vec{\omega}_1 \int \rho_{1m}\rho_{1n} dm_1 
= \frac{1}{2}\,\vec{s}_1\, \delta_{mn}$$ 
of the spin $\vec{s}_1$ of the top 1, it generates
$$\frac{3}{2}\,\frac{k}{c^2 r^3}\,\frac{m_2}{m_1}
\,\vec s_1[\vec r\times \vec p_1]$$
in the spin-orbit potential.
Treating in this way the next term, that with $v_2^2$, in (\ref{L}), we
completely restore the spin-orbit potential (\ref{1ls}) in the
center-of-mass system for the binary, where $\vec p_1 =-\vec p_2=\vec
p$. The similar procedure applied to the terms with
$-\,7(\vec{v}_1\vec{v}_2)-\,(\vec{v}_1\vec{n})(\vec{v}_2\vec{n})$
in (\ref{L}) leads to the next spin-orbit contribution (\ref{2ls}), as
well as to the spin-spin potential (\ref{ss}). 

It should be mentioned that the above expressions for the spin-orbit and
spin-spin interaction in the two-body problem were obtained previously in
Refs. \cite{bg,bo} from the analysis of the scattering amplitude for
spin-1/2 particles in the one-graviton-exchange approximation. (As to
the spin-independent relativistic correction, some terms of this type
are missing from their expression.)  

An amusing fact is that the obtained spin-orbit and spin-spin
interactions are exact analogues (up to an obvious change of notations)
of the corresponding well-known terms in the hydrogen atom. We mean 
the fine and hyperfine structure, the last interaction being induced by
the coupling between the nuclear spin and the electron orbital angular
momentum and spin. (Of course, in our classical approach we cannot
reproduce the contact Fermi spin-spin interaction with $\delta(\vec
r)\,)$.

However, the expression for the spin-orbit correction to the
acceleration, presented in Refs. \cite{kww,bdi}, differs from that which
can be derived from our formulae $V_{1ls}$ and $V_{2ls}$. The discrepancy
is due to the difference in the definitions of the center-of-mass
coordinate of a rotating star. The coordinate $\vec x_i$ advocated and
used in Ref. \cite{kww} is related to ours $\vec r_i$ as follows:
\begin{equation}\label{def}
\vec r_i=\,\vec x_i+\,\frac{1}{2 m_i}\vec v_i\times\vec s_i 
\end{equation}
(from now on we put $c=1$ in our explicit formulae). The shift by itself
is of course a matter of convention, but it is in fact our definition
which {\it just by construction} (see Eq. (\ref{cm}) and the arguments
leading to it) corresponds to the true center-of-mass coordinate of a
rotating star.

Still, what is the meaning of the vector $\vec x$ and why is it
irrelevant to the problem under consideration? The answer can be
conveniently formulated with the following example. For the free Dirac
particle with the Hamiltonian
$$H\,=\,\vec{\alpha}\vec p\, + \,\beta\,m$$ 
the operator whose expectation value equals to $\vec r$, is not $\vec r$
itself, but \cite{fw}
\begin{equation}\label{co}
\vec x\,=\,\vec r\,+\,\frac{i\beta\vec{\alpha}}{2 E_p}\,
-\,\frac{i\beta(\vec{\alpha}\vec p)\vec p\,+\,[\vec{\Sigma}\times
\vec p]\,p}{2 E_p (E_p+m) p};\;\;E_p=\,\sqrt{p^2+m^2};\;\;
\vec{\Sigma}\,=\,\frac{1}{2i}[\vec{\alpha}\times\vec{\alpha}].
\end{equation}
To lowest nonvanishing orfer in $c^{-2}$ expression (\ref{co}) reduces to
\begin{equation}\label{con}
\vec x\,=\,\vec r\,-\,\frac{1}{2 m}\,\vec v\times\vec s;\;\;
\vec s\,=\,\frac{\vec{\sigma}}{2}
\end{equation}
which might prompt indeed substitution (\ref{def}). However, the
transition from the exact Dirac equation in an external field to its
approximate form containing only the first-order correction in $c^{-2}$
is performed by means of the Foldy-Wouthuysen (FW) transformation. And
under the same FW transformation the relativistic operator $\vec x$
(its form for an interacting particle is more complicated than
(\ref{co})) goes over into mere $\vec r$. In other words, in the arising
Hamiltonian the coordinate of spinning electron has the same meaning
$\vec r$ as in the completely nonrelativistic case. Nobody makes
substitution (\ref{def}) when treating the spin-orbit interaction in the
hydrogen atom.

\bigskip

{\bf 3.} Let us consider now the fully covariant equation of motion for a
spinning particle in an external field 
\begin{equation}\label{pa}
\frac{D}{D\tau}\left(m\,u^{\mu}\,
+\,\frac{DS^{\mu\nu}}{D\tau}\,u_{\nu}\right)\,=
-\,\frac{1}{2}R^{\mu\nu}_{\rho\sigma}u_{\nu}S^{\rho\sigma}\,
+\,eF^{\mu\nu}u_{\nu} 
\end{equation}
derived by Papapetrou \cite{pa}. Here $D/D\tau$ means the covariant
derivative with respect to the proper time; $u^{\mu}=dx^{\mu}/dt$ is
the four-velocity; $S^{\mu\nu}$ is the antisymmetric tensor of spin;
$R^{\mu\nu}_{\rho\sigma}$ is the Riemann tensor. We have
included as well into this equation the interaction with an external
electromagnetic field $F^{\mu\nu}$. A close analogy between
the two terms, electromagnetic and gravitational, in the rhs of Eq.
(\ref{pa}) was emphasized in Ref. \cite{kh}.

We will use the common definition of the relativistic spin. According
to it, the only nonvanishing components of the tensor of spin (and the
vector of spin) in the particle rest frame are the space ones.
Transition to an arbitrary frame is performed by a boost. This
definition guarantees automatically the constraint for spin
\begin{equation}\label{cons}
S^{\mu\nu}u_{\nu}=0.
\end{equation}
Due to this constraint,
\begin{equation}\label{id}
\frac{DS^{\mu\nu}}{D\tau}\,u_{\nu}
=\,-\,S^{\mu\nu}\frac{Du_{\nu}}{D\tau}.
\end{equation}
So, if the electromagnetic field is switched off and terms
nonlinear in spin neglected, the second term in the lhs of Eq.
(\ref{pa}) should be deleted. Clearly,  
$-\,\frac{1}{2}R^{\mu\nu}_{\rho\sigma}u_{\nu}S^{\rho\sigma}$ is nothing
else but a covariant expression for the force due to the spin-orbit
interaction. In the field created by a heavy mass $M$ this term
reduces to first order in $c^{-2}$ to
\begin{equation}\label{wa}
-3\,\frac{k M}{r^3}\left(\vec v\times\vec s\,-\,(\vec n \vec v)\,
\vec n\times\vec s\,-\,2\vec n(\vec n[\vec v\times\vec s])\right)
\end{equation}
which coincides with the corresponding force from Ref. \cite{kww}.
However, the force extracted from potential (\ref{1ls1}) is different: 
\begin{equation}\label{ra}
-3\,\frac{k M}{r^3}\left(\vec v\times\vec s\,
-\,\frac{3}{2}\,(\vec n \vec v)\,\vec n\times\vec s\,
-\,\frac{3}{2}\,\vec n(\vec n[\vec v\times\vec s])\right).
\end{equation}
This discrepancy was pointed out long ago in Ref. \cite{bo1} where the
force (\ref{ra}) was derived from the scattering amplitude for the
Dirac particle. The explanation suggested in Ref. \cite{bo1} for the
disagreement is that expression (\ref{wa}) refers to an extended body and
(\ref{ra}) to a point particle. It does not look satisfactory.
For instance, is the proton in a gravitational field a point particle or
extended body? Obviously, as long as we do not go into details of its
structure, as long as we do not consider its internal excitations, an
extended body can be treated as a point particle.

To make the problem even more acute, let us consider another limit,
that of vanishing gravitational field. Here Eq. (\ref{pa}) describes a
particle with spin, but without magnetic moment. Still, its spin
interacts with an external electromagnetic field, which to lowest order
in $c^{-2}$ should be described by the Thomas interaction 
\begin{equation}\label{th}
V_t=\,\frac{e}{2 m}\,\vec s\,[\vec E\times\vec v].
\end{equation}
This expression can be easily recovered from the well-known results for
the spin precession (see, e.g., book \cite{blp}) at the vanishing
$g$-factor, $g=0$. When the electric field $\vec E$ is that of a point
charge $-Ze$, the force corresponding to interaction (\ref{th}) is
\begin{equation}\label{rat}
\frac{Ze^2}{r^3}\left(\vec v\times\vec s\,
-\,\frac{3}{2}\,(\vec n \vec v)\,\vec n\times\vec s\,
-\,\frac{3}{2}\,\vec n(\vec n[\vec v\times\vec s])\right).
\end{equation}
However, the force obtained in this case from the second term in the lhs
of Eq. (\ref{pa}) to lowest order in $c^{-2}$, is different:
\begin{equation}\label{wat}
\frac{Ze^2}{r^3}\left(\vec v\times\vec s\,-\,(\vec n \vec v)\,
\vec n\times\vec s\,-\,2\vec n(\vec n[\vec v\times\vec s])\right).
\end{equation}

The reason of both discrepancies is clear now: $\vec r$ entering
expressions (\ref{wa}), (\ref{wat}) is just the relativistic coordinate
of Eq. (\ref{pa}), it is nothing else but $\vec x$ in the notations
of relations (\ref{co}). Therefore, the transition from the
fully relativistic Eq. (\ref{pa}) to the $c^{-2}$ approximations to it
(\ref{wa}), (\ref{wat}), should be accompanied indeed by substitution
(\ref{con}). This substitution should be performed of course both in the
acceleration entering the Newton equation of motion, and in the
(formally) nonrelativistic force. In this way correct Eqs. (\ref{ra}),
(\ref{rat}) are restored.

\bigskip

{\bf 4.} The above consideration of the Papapetrou equation (\ref{pa})
is instructive in one more respect. It was pointed out that
this equation describes a particle with spin, but without magnetic
moment. The magnetic moment interaction is well-known to be taken into
account by the following term in the relativistic Hamiltonian:
\begin{equation}\label{mm}
V_{mm}=\,\frac{e g}{4 m}S^{\mu\nu}F_{\mu\nu}.
\end{equation}
In Ref. \cite{kh} it was demonstrated that expression (\ref{mm}) has a
close gravitational analogue
\begin{equation}\label{gm}
V_{gm}=\,-\,\frac{\kappa}{8 m}S^{\mu\nu}S^{\rho\sigma}
R_{\mu\nu\rho\sigma}
\end{equation}
which can be called gravitational magnetic moment interaction.
In particular, this coupling arises in a natural way in wave
equations, and the value $\kappa = 1$ for the constant in it is as
preferable from the point of view of the high-energy behaviour as $g=2$
is in the electromagnetic case \cite{kh}. Both interactions, (\ref{mm})
and (\ref{gm}), should include in fact some additional terms treated in
detail in Refs. \cite{fr} (for usual magnetic moment) and \cite{yb}.
Being certainly of higher order in $v/c$, those terms can be omitted in
our treatment of binary stars. 

For the field created by a heavy mass $M$ interaction (\ref{gm})
reduces in lowest, first order in $c^{-2}$ to the quadrupole form:
\begin{equation}\label{1s}
V^1_{s}=\,\frac{3 k M}{2 r^3}Q_{mn}^s n_m n_n
\end{equation}
where the effective quadrupole moment
$$Q^s_{mn}=\,\frac{1}{m}\,(\,s_m s_n-\,\frac{1}{3}\,\delta_{mn}s^2)$$
resembles by its spin dependence the well-known expression from quantum
mechanics. For the two-body problem under discussion expression
(\ref{1s}) generalizes to the following self-interaction of spins:
\begin{equation}\label{s}
V_{s}=\,\kappa\,\frac{k}{2 r^3}\left(\frac{m_2}{m_1}s_{1m} s_{1n}
+\,\frac{m_1}{m_2}s_{2m} s_{2n}\right)\,(3n_m n_n-\,\delta_{mn}).
\end{equation}
resembling the usual spin-spin interaction (\ref{ss}). 

Let us compare now the effective quadrupole interaction (\ref{1s}) or
(\ref{s})  with the usual quadrupole one.  At $\kappa\, \sim\, 1$
interaction (\ref{s}) is of the same order of magnitude as the spin-spin
one (\ref{ss}). Even in the most favourable case when they can become
essential, that of two extreme Kerr black holes, both interactions are
of the $c^{-4}$ order. The star rotation velocity is here $\sim\,c$, but
its radius is close to the gravitational one $r_g\sim c^{-2}$, so that
each spin $s\,\sim c^{-1}$ \cite{kww}. (The same argument demonstrates
that the spin-orbit interaction is of the $c^{-3}$ order \cite{kww}.) As
to the usual quadrupole interaction, it is suppressed by the small value
of the quadrupole deformation and, according to Ref. \cite{bc}, can also
manifest itself in the case of two extreme Kerr black holes only. 

\bigskip

{\bf 5.} We are going over at last to the spin effects in the
gravitational radiation of binary stars. In fact, the only essentially
new correction to the energy loss obtained by us is that due to the spin
self-interaction and originating mainly from interaction (\ref{s}). Our
final result for the contribution due to the spin-spin interaction
(\ref{ss}) coincides with that presented in Refs. \cite{kww,bdi}. As to
the spin-orbit correction, our result for it can be in fact obtained
from the expression given in \cite{kww,bdi} by merely going back to the
simple-minded definition of the center-of-mass coordinate advocated by
us above. However, in parallel with calculating the correction due to
the spin self-interaction (\ref{s}), we will present corresponding
contributions induced by the spin-orbit and spin-spin couplings
(\ref{1ls}), (\ref{2ls}), (\ref{ss}). It serves as an independent check
of the results presented previously in Refs. \cite{kww,bdi} (an
arithmetical confirmation only in the case of the spin-orbit
interaction).

We start with the well-known expression for the metric perturbation
$h_{mn}$ at large distance $R$ from the source (see, e.g., \cite{ll}, 
\S 110):
\begin {equation}\label{rad}
\psi_{mn}(R,t)\,=\,-\,\frac{4 k}{R}
\int d\vec r \tau_{mn}(\vec r,\,t-R+\vec r \vec n);\;\;\;
\psi_{mn}=\,h_{mn}+\,\frac{1}{2}\,\delta_{mn}h_{\rho\rho};\;\;\;
\vec n\,=\,\frac{\vec R}{R}.
\end{equation}
The source $\tau_{mn}$ includes not only the energy-momentum
tensor of matter, but generally speaking corresponding nonlinearities
of the gravitational field itself. It is conserved in the sense
\begin {equation}\label{cons}
\partial_{\mu}\tau_{\mu\nu}=\,0.
\end{equation}
As usual, we will be interested in the part of the perturbation
$\psi_{mn}$ which is orthogonal to
$\vec n$ and trace-free (otherwise Eq. (\ref{rad}) would look
slightly more complicated). It should be mentioned here that both
expression (\ref{rad}) and the
$c^{-2}$ Lagrangian (\ref{L}) are valid under the same gauge conditions
$$\partial_m h_{m0}-\,\frac{1}{2}\partial_0 h_{mm}=\,0,\;\;\;
\partial_{\mu} h_{\mu n}-\,\frac{1}{2}\partial_n h_{\mu\mu}=\,0.$$
One can easily check it by inspecting the corresponding derivations in
book \cite{ll} (\,\S\S 106, 110\,). 

Neglecting the retardation $\vec r \vec n$ in expression (\ref{rad})
we reduce it to the quadrupole formula
\begin{equation}
\psi_{mn}^0\,=\,-\,\frac{2 k}{R}\,\partial^2_0 
\int d\vec r\,r_m r_n \tau_{00}.
\end{equation}
To lowest order in $v/c\;\; \tau_{00}$ reduces to rest masses, and the
integral gives the usual quadrupole moment 
$$Q_{mn}=\,\mu\,(\,r_m r_n-\frac{1}{3}\,\delta_{mn});\;\;\;
\mu\,=\,\frac{m_1 m_2}{m_1+\,m_2}.$$  
Here new
terms in the quadrupole radiation intensity are generated by the
spin-dependent corrections to the orbit radius $r$ and to the equations
of motion used to evaluate time derivatives of $\vec r$. In all our
discussions of gravitational radiation we restrict to the case of
circular orbits which is most interesting from the physical point of
view \cite{kww}, and much more simple as concerns calculations. In this
way we get the following relative corrections to the usual quadrupole
formula:
\begin{equation}\label{q1ls}
\frac{I^{q1}_{ls}}{I^q}=\,-\,\frac{1}{m_1 m_2 r^2}\,\vec l\, 
(\,6\,\vec s\,+\,\frac{9}{2}\vec{\xi}\,);
\end{equation}
\begin{equation}\label{q1ss}
\frac{I^{q1}_{ss}}{I^q}=\,\frac{9}{2 m_1 m_2 r^2}\, 
(\,3\,s_{1t}s_{2t}-\,\vec s_1\vec s_2);
\end{equation}
\begin{equation}\label{q1s}
\frac{I^{q1}_{s}}{I^q}=\,\frac{9\kappa}{4 m_1 m_2 r^2}\, 
\left[\,\frac{m_2}{m_1}\,(3\,s_{1t}^2-\,s_1^2)\,+\,
\frac{m_1}{m_2}\,(3\,s_{2t}^2-\,s_2^2)\,\right].
\end{equation}
Here
$$I^q=\,\frac{32 k^4 m_1^2 m_2^2 m}{5 r^5}$$
is the unperturbed quadrupole intensity, $m=\,m_1+\,m_2$; subscripts
{\it ls, ss, s} refer to the spin-orbit, spin-spin and
spin-self-interaction contributions respectively; 
$$\vec s=\vec s_1+\vec s_2;\;\;
\vec{\xi}=\,\frac{m_2}{m_1}\,\vec s_1+\,\frac{m_1}{m_2}\,\vec s_2.$$
The expressions for $I^{q1}_{ss}$ and $I^{q1}_{s}$ have been averaged
over the period of rotation. That is why they contain the spin
components $s_{t}$ orthogonal to the orbit plane. 

The next correction to the qudrupole radiation originates from the
terms of the relative order $c^{-2}$ in $\tau_{00}$. The only
spin-dependent contribution here is of the $ls$ type. The same procedure
which has generated the spin interactions from Lagrangian (\ref{L})
allows one to extract from
$$\frac{m_1 v^2_1}{2}\,+\,\frac{m_2 v^2_2}{2}$$
the following correction to the quadrupole moment
\begin{equation}\label{de1}
\delta Q_{mn}^1=\,\frac{\mu}{m}\,r_m \epsilon_{nrs} v_r \xi_s.
\end{equation}
Since this expression will
be anyway contracted with the symmetric $Q_{mn}$, there is no need to
symmetrize it explicitly. Certainly, correction (\ref{de1}) makes a
spin-orbit type contribution to the radiation intensity only. But we
will postpone its calculation for the time being.

Let us go over now to the retardation effects in radiation. The
first-order correction in formula (\ref{rad}) looks as follows:
\begin{equation}
\psi_{ab}^1\,=\,-\,\frac{4 k}{R}\,\partial_0 
\int d\vec r\,z\,\tau_{ab}.
\end{equation}
We have made explicit here that the wave propagates along the $z$ axis,
$a,b\,=\,1,2$. Simple transformations based on the continuity equation
(\ref{cons}) (see \cite{ll}, \S 110) lead to the following identity:
\begin{equation}\label{id}
\int d\vec r\,r_k\tau_{mn}=\,\frac{1}{6}\partial_0^2\int d\vec r\,
r_k r_m r_n \tau_{00}+\,\frac{2}{3}\,\partial_0\,\int d\vec
r\,r_n (r_k \tau_{0m}-\,r_m \tau_{0k}).
\end{equation}
The first, totally symmetric term in this expression generates the
octupole radiation. Being spin-independent, it is not of interest to
us. 

In the second structure we restrict to the term in $\tau_{0m}$
which is of lowest order in $v/c$, and obtain in this way
\begin{equation}
\int d\vec r\,r_n (r_k \tau_{0m}-\,r_m \tau_{0k})\,=
\,m_1(r_{1m} r_{1n} v_{1k}-\,r_{1m} r_{1n} v_{1k})\,
+\,m_2(r_{2m} r_{2n} v_{2k}-\,r_{2m} r_{2n} v_{2k}).
\end{equation}
With the previous trick we single out in this tensor the spin-dependent
terms and arrive at the expression which can be presented as
$$J_{kmn}=\,\epsilon_{nkl}J_{lm}.$$
The second-rank tensor here is
\begin{equation}
J_{mn}=\,-\,\frac{m_1-\,m_2}{2 m}\,(\,r_{m}l_{n}+\,r_{n} l_{m})
+\,\frac{3}{4}\,\mu\,(\,r_m \zeta_n +\,r_n \zeta_m 
-\,\frac{2}{3}\,\vec r \vec{\zeta}\,)
\end{equation}
where
$$\vec{\zeta}\,=\,\frac{\vec s_1}{m_1}\,-\,\frac{\vec s_2}{m_2}.$$
It is a close analogue of the magnetic quadrupole moment in
electrodynamics, one can single out in it in an obvious way the
contributions of the convection and spin currents.

The intensity of this gravimagnetic quadrupole radiation can be
conveniently calculated via the following transformation of the initial
structure $n_k e_{mn} J_{kmn}$: 
\begin{equation}
n_k e_{mn} J_{kmn}=\,\epsilon_{lnk} e_{mn} n_k J_{lm}
=\,\tilde{e}_{lm}J_{lm}.
\end{equation}
If we choose the independent components of the polarization tensor as
$$e_{mn}=\,\left(\,\frac{e_{11}-\,e_{22}}{2},
\;\frac{e_{12}+\,e_{21}}{2}\,\right),$$
then the dual polarization is
\begin{equation}
\tilde{e}_{lm}=\,\epsilon_{lnk} e_{mn} n_k
=\,\left(\,-\frac{e_{12}+\,e_{21}}{2},\;\frac{e_{11}-\,e_{22}}{2}\,\right).
\end{equation}
Formally the sum over independent dual polarizations $\tilde{e}$ in
$(\,\tilde{e}_{lm}J_{lm})^2$ looks exactly the same as that
over common polarizations $e$ when calculating the usual quadrupole
radiation. In the present case the intensity equals \cite{kww,th}
\begin{equation}\label{gmq}
I^{gmq}=\,\frac{16}{45}\,k^4\,J^{(3)}_{mn}J^{(3)}_{mn}
\end{equation}
where the superscript at $J$ denotes the third time derivative. The
calculation of these derivatives is simplified in the present case by
the fact that to our accuracy both $\vec l$ and $\vec{s}_i$ can be
considered as constant in time. The spin-dependent corrections arising in
this way are
\begin{equation}\label{gmqls}
\frac{I^{gmq}_{ls}}{I^q}=\,-\,\frac{1}{12\,m_1 m_2 r^2}\,\vec l\, 
(\,\vec s\,-\,\vec{\xi}\,);
\end{equation}
\begin{equation}\label{gmqss}
\frac{I^{gmq}_{ss}}{I^q}=\,\frac{1}{48\,m_1 m_2 r^2}\, 
(\,s_{1t}s_{2t}-\,7\,\vec s_1\vec s_2).
\end{equation}
Even a spin-self-interaction correction (somehow missed in Ref.
\cite{kww}) arises here:
\begin{equation}\label{gmqs}
\frac{I^{gmq}_{s}}{I^q}=\,-\,\frac{1}{96\,m_1 m_2 r^2}\, 
\left[\,\frac{m_2}{m_1}\,(\,s_{1t}^2\,-\,7\,s_1^2)\,+\,
\frac{m_1}{m_2}\,(\,s_{2t}^2\,-\,7\,s_2^2)\,\right].
\end{equation}
 
Let us consider at last the second-order retardation correction
in formula (\ref{rad}) 
\begin{equation}
\psi_{ab}^2\,=\,-\,\frac{4 k}{R}\,\frac{1}{2}\,\partial_0^2 
\int d\vec r\,z^2\,\tau_{ab}.
\end{equation}
A spin contribution can be produced in it only by velocity-dependent
terms in $r_k r_l\tau_{mn}$, in other words only by the energy-momentum
tensor of matter. These terms are of the type $v_m v_n r_k r_l$ and
generate the following structure
$$2\,\frac{\mu}{m}\,r_k v_m\epsilon_{lnr}\xi_r.$$
The symmetrizations $(\,k,\, l\,)$ and $(\,m,\,n\,)$ are implied here.
The irreducible part of the third-rank tensor $r_k v_m \xi_r$ can be
omitted at once since it does not interfere in the total intensity with
the second-rank tensor $Q_{mn}$. Then, we omit also the structures of
the type $(\,r_k v_m-\,r_m v_k)\,\xi_r$ since
both orbital angular momentum and spin can be considered constant
in time to our accuracy. The resulting structure transforms as
follows:
$$\frac{2}{3}\,\frac{\mu}{m}\,n_k n_l e_{mn} \epsilon_{lnr} v_m
\,(\,r_k\xi_r-\,r_r\xi_k)\,
=\,\frac{2}{3}\,\frac{\mu}{m}\,n_k n_l e_{mn} \epsilon_{lnr} v_m
\epsilon_{krs}\epsilon_{ijs} r_i \xi_j
=\,-\,\frac{2}{3}\,\frac{\mu}{m}\,e_{mn} v_m
\epsilon_{nrs} r_r \xi_s.$$
In other words, this correction to the quadrupole moment is
\begin{equation}\label{de2}
\delta Q_{mn}^2=\,-\,\frac{2}{3}\,\frac{\mu}{m}\,v_m \epsilon_{nrs} r_r
\xi_s.
\end{equation}
Adding up expressions (\ref{de1}) and (\ref{de2}) and neglecting again
the term $(\,r_m v_r-\,r_r v_m)\,\xi_s$, we obtain the following total
spin-dependent correction to the quadrupole moment:
\begin{equation}\label{de}
\delta Q_{mn}=\,\frac{1}{3}\,\frac{\mu}{m}\,v_m \epsilon_{nrs} r_r
\xi_s.
\end{equation}
The corresponding relative correction to the radiation intensity
constitutes
\begin{equation}\label{q2ls}
\frac{I^{q2}_{ls}}{I^q}=\,\frac{2}{3\,m_1 m_2 r^2}\,\vec l\, \vec{\xi}.
\end{equation}

Now, expressions (\ref{q1ls}), (\ref{gmqls}) and (\ref{q2ls}), taken
together, give the following total spin-orbit correction:
\begin{equation}\label{cls}
\frac{I_{ls}}{I^q}=\,-\,\frac{\vec l\,(\,73\,\vec
s\,+\,45\,\vec{\xi}\,)}{12\,m_1 m_2 r^2}.
\end{equation}
It can be easily checked that the corresponding result of Refs.
\cite{kww,bdi} would be reconciled with this one under the proper
definition of the center-of-mass coordinate.

Adding up expressions (\ref{q1ss}) and (\ref{gmqss}), we obtain the
result of Refs. \cite{kww,bdi} for the spin-spin correction:
\begin{equation}\label{css}
\frac{I_{ss}}{I^q}=\,\frac{1}{48\,m_1 m_2 r^2}\, 
(\,649\,s_{1t}s_{2t}-\,223\,\vec s_1\vec s_2).
\end{equation}

And at last, the total spin-self-interaction correction, generated by
(\ref{q1s}) and (\ref{gmqs}), is
\begin{equation}
\frac{I_{s}}{I^q}=\,\frac{1}{4 m_1 m_2 r^2}\, 
\left[\,(\,27\,\kappa\,-\,\frac{1}{24}\,)\,
(\,\frac{m_2}{m_1}\,s_{1t}^2\,+\,
\frac{m_1}{m_2}\,s_{2t}^2\,)\,-\,
(\,9\,\kappa\,-\,\frac{7}{24}\,)\,(\,\frac{m_2}{m_1}\,s_1^2\,+\,
\frac{m_1}{m_2}\,s_2^2)\,\right].
\end{equation}
As has been mentioned already, at $\kappa\,\sim\,1$ this new correction
is quite comparable to the spin-spin one.

\bigskip
\bigskip
\bigskip

We are grateful to A.I. Vainshtein for the discussion of
results and reading the manuscript. We are also grateful to N.A. Voronov
and A.F. Zakharov for their interest to the work. One of us
(I.B. Kh.) acknowledges support by the Russian Foundation for Basic
Research through grant No.95-02-04436-a, by the Program "Universities of
Russia" through grant No. 94-6.7-2053, and by the National Science
Foundation through a grant to the Institute for Theoretical Atomic and
Molecular Physics at Harvard University and Smithsonian Astrophysical
Observatory.

\pagebreak

\end{document}